\documentclass[trackchanges,twocolumn,tighten]{aastex62}

\usepackage{amsmath}
\usepackage{longtable}

\newcommand{\dif}{\ensuremath{\mathrm{d}}}

\newcommand{\code}[1]{\texttt{#1}}
\newcommand{\mesa}{\code{MESA}}
\newcommand{\MESA}{\mesa}

\newcommand{\stella}{\code{STELLA}}

\newcommand{\STELLA}{\stella}

\newcommand{\Lsun}{L_\odot}
\newcommand{\Msun}{M_\odot}
\newcommand{\Rsun}{R_\odot}
\newcommand{\Mej}{M_\mathrm{ej}}

\newcommand{\MZAMS}{M_\mathrm{ZAMS}}
\newcommand{\Mzams}{\MZAMS}

\newcommand{\Mf}{M_\mathrm{final}}
\newcommand{\Mten}{M_{10}}

\newcommand{\Rfh}{R_{500}}
\newcommand{\Eexp}{E_\mathrm{exp}}

\newcommand{\Efoe}{E_{51}}
\newcommand{\Lfifty}{L_{50}}
\newcommand{\Lft}{L_{42}}
\newcommand{\vfifty}{v_{50}}
\newcommand{\Teff}{T_\mathrm{eff}}

\newcommand{\Lbol}{L_\mathrm{bol}}

\newcommand{\tCo}{\tau_\mathrm{Co}}

\newcommand{\tni}{t_\mathrm{Ni}}
\newcommand{\tNi}{\tni}

\newcommand{\tpt}{t_\mathrm{p}}

\newcommand{\vFef}{v_\mathrm{Fe,50}}

\newcommand{\ooc}{\omega/\omega_\mathrm{crit}}

\newcommand{\Ni}{{^{56}\mathrm{Ni}}}
\newcommand{\Co}{{^{56}\mathrm{Co}}}

\newcommand{\MNi}{M_\mathrm{Ni}}
\newcommand{\foe}{10^{51}\,\mathrm{ergs}}

\newcommand{\ET}{ET}

\newcommand{\etaNi}{\eta_\mathrm{Ni}}
\newcommand{\QNi}{Q_\mathrm{Ni}}

\newcommand{\tptwo}{t_\mathrm{p, 2}}

\newcommand{\days}{\mathrm{d}}

\newcommand{\appropto}{\mathrel{\vcenter{
		\offinterlineskip\halign{\hfil$##$\cr
	\propto\cr\noalign{\kern2pt}\sim\cr\noalign{\kern-2pt}}}}}

\newcommand{\gtapprox}{\mathrel{\vcenter{
		\offinterlineskip\halign{\hfil$##$\cr
	>\cr\noalign{\kern2pt}\sim\cr\noalign{\kern-2pt}}}}}
\newcommand{\ltapprox}{\mathrel{\vcenter{
		\offinterlineskip\halign{\hfil$##$\cr
	<\cr\noalign{\kern2pt}\sim\cr\noalign{\kern-2pt}}}}}

\newcommand{\mesafour}{MESA~IV} 

\newlength{\apjcolwidth}
\newlength{\figwidth}
\newlength{\doublewide}
\begin{document}
\title{The Value of Progenitor Radius Measurements for Explosion Modeling of Type II-Plateau Supernovae} 

\author[0000-0003-1012-3031]{Jared A. Goldberg}
\affiliation{Department of Physics, University of California, Santa Barbara, CA 93106, USA}

\author{Lars Bildsten}
\affiliation{Department of Physics, University of California, Santa Barbara, CA 93106, USA}
\affiliation{Kavli Institute for Theoretical Physics, University of California, Santa Barbara, CA 93106, USA}

\correspondingauthor{J. A. Goldberg}
\email{goldberg@physics.ucsb.edu}

\begin{abstract}
Using Modules for Experiments in Stellar Astrophysics (\MESA)+\STELLA, we show that very different physical 
models can adequately reproduce a specific observed Type II-Plateau Supernova (SN).
We consider SN2004A, SN2004et, SN2009ib, SN2017eaw, and SN2017gmr, 
Nickel-rich ($M_\mathrm{Ni}>0.03M_\odot$) events with bolometric lightcurves and a well-sampled decline from the plateau. These
events also have constraints on the progenitor radius, via a progenitor image, or, in the case of SN2017gmr, 
a radius from fitting shock-cooling models. In general, many explosions spanning the parameter space of progenitors 
can yield excellent lightcurve and Fe line velocity agreement, demonstrating the success of scaling laws in motivating 
models which match plateau properties for a given radius and highlighting the degeneracy between plateau luminosity and 
velocity in models and observed events, which can span over 50\% in ejecta mass, radius, and explosion energy. 
This can help explain disagreements in explosion properties reported for the same event using different model calculations. 
Our calculations yield explosion properties when~combined with pre-explosion~progenitor radius measurements 
or a robust understanding of the outermost $<0.1\,M_\odot$ of material that 
quantifies the progenitor radius from SN observations a few days after explosion.
\end{abstract}

\keywords{
hydrodynamics --- radiative transfer --- stars: massive --- supernovae: general --- supernovae: individual (2004A, 2004et, 2009ib, 2017eaw, 2017gmr)
}


\section{Introduction\label{sec:INTRODUCTION}}

Massive stars ($M\gtapprox10\Msun$) at the end of their evolution become
red supergiants (RSGs) with radii of $\approx400-1000\Rsun$, 
before ending their lives as core-collapse Type IIP supernovae (SNe) with
lightcurves that plateau over $\approx100$ days.
The progenitor radius $(R)$, ejected mass ($\Mej$), explosion energy ($\Eexp$), and $\Ni$ mass ($\MNi$)
determine these lightcurves \citep[e.g.][]{Popov1993,Sukhbold2016}, and inferring these properties 
from observations could lend insight into which stars explode as SNe. 
Although early work provided scaling relations attempting to uniquely relate plateau properties and expansion 
velocities to explosion characteristics \citep[e.g.][]{Litvinova1983,Popov1993},
recent work highlights the nonuniqueness of 
lightcurve and velocity modeling for a given SN after $\approx$20 days \citep{Dessart2019,Goldberg2019,Martinez2019}. 

Building on \citet[hereafter GBP19]{Goldberg2019}, we 
verify these degeneracies by comparing explosions of very different progenitor models to 
Nickel-rich ($M_\mathrm{Ni}>0.03M_\odot$) events with bolometric lightcurves,
a well-sampled decline from the plateau, and constraints on the progenitor radius.
We utilize the open-source 1D stellar evolution code Modules for Experiments in Stellar Astrophysics
\citep[\MESA,][]{Paxton2011, Paxton2013, Paxton2015, Paxton2018, Paxton2019} for our evolutionary and explosion 
models and the multi-group radiation-hydrodynamics instrument 
\STELLA\ \citep{Blinnikov1998, Blinnikov2000, Blinnikov2006} to produce 
lightcurves and model expansion velocities. 
Emission in the first 20 days depends on the radial density
structure of the outer $<0.1\Msun$ of matter around a vigorously convecting RSG
progenitor \citep[e.g.][]{Morozova2016}. SN emission during this time
can be modified by the uncertain circumstellar environment (e.g.\citealt{Morozova2017}), and
may reflect the intrinsically 3D structure of these outer layers (see e.g. \citealt{Chiavassa2011}).
Therefore we restrict our analysis to observations after day $\approx$20, when emission 
comes from the bulk of the stellar envelope. However, we still show our results for earlier 
times, where the qualitative trends may hold.

\section{Observed Supernovae and Their Degeneracy Curves\label{sec:degeneracy}}

GBP19 showed that Type IIP supernovae with $\Ni$ mass 
($\MNi\geq0.03\Msun$), luminosity at day 50 $(\Lfifty)$, and plateau duration $(\tpt)$ can approximately yield
the ejected mass $(\Mten\equiv\Mej/10\Msun)$ and asymptotic explosion energy $(\Efoe\equiv\Eexp/10^{51}\mathrm{ergs})$ 
as a function of progenitor radius ($\Rfh\equiv R/500\Rsun$), 
via the following relations:
\begin{equation}
\begin{split}
\begin{aligned}
\log(\Efoe)&=-0.728+2.148\log(\Lft)-0.280\log(\MNi)\\&+2.091\log(\tptwo)-1.632\log(\Rfh),\\
\log(\Mten)&=-0.947+1.474\log(\Lft)-0.518\log(\MNi)\\&+3.867\log(\tptwo)-1.120\log(\Rfh),
\label{eq:scaling}
\end{aligned}
\end{split}
\end{equation}
where $\MNi$ is in units of $\Msun$, $\Lft=\Lfifty/10^{42}$ erg s$^{-1}$ 
and $\tptwo=\tpt/100\,\days$, and log is base 10. 
Moreover, because expansion velocities inferred from the Fe II 5169\AA\ line are determined by 
line-forming regions near the photosphere, velocity data during the 
plateau period do not break this degeneracy ($\Lfifty\appropto\vfifty^2$, \citealt{Hamuy2002,Kasen2009}). 
Rather, SNe with the same $\Lfifty$, $\tpt$, and $\MNi$ and similar expansion velocities during the plateau
can be realized by a family of explosions with a range of $R$, $\Eexp$, and $\Mej$
obeying the Equation~\eqref{eq:scaling} relations. 

\subsection{Measuring Nickel Mass and Plateau Duration of Type IIP SNe}
We estimate the plateau duration $\tpt$ following \citet{Valenti2016}, 
fitting the functional form $y(t)$ to the bolometric luminosity ($\Lbol$) around 
the fall from the plateau:
\begin{equation}
y(t)\equiv\log(\Lbol)=\frac{-A_0}{1+e^{(t-\tpt)/W_0}}+(P_0\times t)+M_0.
\label{eq:FermiFunction}
\end{equation}
We use the python routine \texttt{scipy.optimize.curve\_fit} to fit the lightcurve starting when the 
luminosity evolution is 75\% of the way to its steepest descent, fixing $P_0$ to be the 
slope on the $\Ni$ tail (GBP19). The fitting parameter $\tpt$ is the plateau duration. 
We also extract the $\Ni$ mass from $\Lbol$ by calculating the cumulative observable 
$\ET$ \citep{Nakar2016}, which corresponds 
to the total time-weighted energy radiated 
away in the SN generated by the initial shock and not by $\Ni$ decay:
\begin{equation}
\ET_\mathrm{c}(t)=\int_{0}^{t} t'\left[\Lbol(t')-\QNi(t')\right]\,\dif t',
\label{eq:ETdef}
\end{equation}
where $t$ is the time in days since the explosion and
\begin{equation}
\QNi=\frac{\MNi}{\Msun}\left(6.45e^{-t'/\tNi}+1.45e^{-t'/\tCo}\right)\times10^{43}\ \mathrm{erg\ s^{-1}},
\label{eq:QNi}
\end{equation}
is the $\Ni\rightarrow\Co\rightarrow\mathrm{^{56}Fe}$ decay luminosity given by \citet{Nadyozhin1994}, 
equivalent to the heating rate of the ejecta assuming complete trapping 
with $\tNi=8.8$ days and $\tCo=111.3$ days.
As $t\rightarrow\infty$ and all shock energy has radiated away, the 
slope of the $\ET_\mathrm{c}$ curve on the $\Co$ decay tail should be zero
when the estimate of $\MNi$ is correct. 
This method yields excellent agreement 
between the resulting model lightcurve tails and observed lightcurves, 
and with
the $\Co$ decay 
luminosity \citep{Nadyozhin1994}:
\begin{equation}
L(t\rightarrow\infty)=1.45\times10^{43}\exp\left(-\frac{t}{\tCo}\right)\frac{\MNi}{\Msun}\ \mathrm{erg\ s^{-1}}.
\end{equation}

\subsection{Supernova Selection}
In order to further explore this degeneracy, we apply these scalings to five observed 
supernovae: \textbf{SN2004A}, \textbf{SN2004et}, \textbf{SN2009ib}, \textbf{SN2017eaw}, and \textbf{SN2017gmr}.

\textbf{SN2004A} was discovered by K. Itagaki on 9 January 2004 in NGC6207 \citep{Hendry2006}. 
Following \citet{Pejcha2015a} we adopt an explosion date of MJD 53001.53. Progenitor observations 
indicate $\log(L_\mathrm{p}/\Lsun)=4.9\pm\,0.3$ and $\Teff=3890\pm\,375\,\mathrm{K}$, 
implying a radius of $\approx625\Rsun$ \citep{Smartt2015}. From the \citet{Pejcha2015a} bolometric lightcurve, 
we get $\log(\Lft)=-0.07$.
Estimates for the $\Ni$ mass include $\MNi/\Msun=0.050^{+0.040}_{-0.020}$ 
from points on the bolometric-corrected V-band tail and 
$\MNi/\Msun=0.042^{+0.017}_{-0.013}$ comparing to the tail of 1987A, 
which the original authors average to yield $\MNi/\Msun=0.046^{+0.0031}_{-0.017}$ \citep{Hendry2006}.
We measure a plateau duration of $\tpt=$124 days and use $\MNi=0.042\Msun$. 

\textbf{SN2004et} was discovered in NGC6946 by S. Moretti on 2004 September 27, 
with a well-constrained explosion date of 22.0 September 2004 (MJD 53270.0)
\citep{Li2005}.  
There is some disagreement in the literature about the progenitor (see \citealt{Smartt2009} and \citealt{Davies2018})
since follow-up imaging show R- and I-band flux excesses 
in the location of the inferred progenitor in HST pre-imaging \citep{Crockett2011}.
As a result, \citet{Martinez2019} report a progenitor radius of $350\Rsun-980\Rsun$.
We adopt the bolometric lightcurve given by \citet{Martinez2019}, which indicates $\log(\Lft)=0.27$. 
Estimates for the $\Ni$ mass include $\MNi/\Msun=0.048\pm0.01$ from the scaled $\Co$ decay tail of 1987A
to $\MNi=0.06\pm0.02$ estimated using V-magnitudes from 250-315 days \citep{Sahu2006}. 
We measure $\tpt=123.1$ days and use $\MNi=0.063\Msun$. 

\textbf{SN2009ib} 
was discovered by the Chilean Automatic Supernova Search
on 6.30 August 2009 in NGC1559, with an estimated explosion date of MJD 55041.3 \citep{Takats2015}. 
HST pre-images indicate either a yellow source
with $\log(L_\mathrm{p}/\Lsun)=5.04\pm0.2$, or possibly a fainter RSG with 
$\log(L_\mathrm{p}/\Lsun)=5.12\pm0.14$ and $R\approx1000\Rsun$ assuming $\Teff\approx3400$K \citep{Takats2015}. 
This event is peculiar in that there is a shallow drop from the plateau luminosity to the $\Co$ decay tail, 
falling noticeably off of the \citet{Muller2017} relation between $\Lfifty$ and $\MNi$. 
From the \citet{Takats2015} lightcurve, 
we measure $\log(\Lft)=-0.33$ and $\MNi/\Msun=0.043$, and $\tpt=139.8,\mathrm{days}$. 
\citet{Nakar2016} also highlighted that this event
had a ratio of the integrated $\Ni$ decay chain energy to integrated shock energy 
of $\etaNi=2.6$, much larger than typical values of $\etaNi\approx0.2-0.6$ 
(e.g. $\etaNi$ for $\mathrm{SN1999em}\approx0.54$). 

\textbf{SN2017eaw} was discovered by P. Wiggins on 14.238 May 2017 in NGC6946, 
with an estimated explosion date of MJD 57886.0 \citep{Szalai2019}. 
Pre-explosion imaging suggests$\log(L_\mathrm{p}/\Lsun)=4.9\pm0.2$ 
and $\Teff=3350^{+450}_{-250}$ K, corresponding to $R\approx845\Rsun$,
obscured by a $>2\times10^{-5} \Msun$ dust shell extending out to $4000\Rsun$ 
\citep{Kilpatrick2018}, assuming the distance to NGC6946 to be $D=6.72\pm0.15\,\mathrm{Mpc}$  
(from the tip of the red giant branch (TRGB) by \citealt{Tikhonov2014}).
We adopt the bolometric lightcurve of \citet{Szalai2019} using $D=6.85$~Mpc,
although more recent TRGB measurements suggest $D=7.72\pm0.78$ Mpc \citep{VanDyk2019}.
Estimates for the $\Ni$ mass assuming $D=6.85$~Mpc range from $\MNi/\Msun=0.036-0.045$ 
\citep{Szalai2019} to $\MNi=0.05\Msun$ \citep{Tsvetkov2018}. 
From the \citet{Szalai2019} lightcurve, we measure $\tpt=117.2$ days, 
$\MNi=0.048\Msun$, and $\log(\Lft)=0.21$.

\textbf{SN2017gmr} occurred in NGC988, discovered on MJD 58000.266 during the
DLT40 SN search with the explosion epoch assumed to be MJD 57999.09 at $D=19.6\pm$1.4 Mpc 
\citep{Andrews2019}.
No progenitor detection was made, but shock-cooling modeling of the early SN
recovers $R\approx500\Rsun$. \citet{Andrews2019} find $\MNi=0.13\pm0.026\Msun$ assuming 
all late-time luminosty comes from Ni decay, although multipeaked emission lines emerging after day 150 suggest 
asymmetries present either in the core's explosion or in late-time interaction with the surrounding environment. 
We adopt the \citet{Andrews2019} bolometric lightcurve, 
and measure $\log(\Lft)=0.57$, $\MNi/\Msun=0.13$, and $\tpt=$94.5 days. 

\subsection{The Degeneracy Curves}
\begin{figure}
\centering
\includegraphics{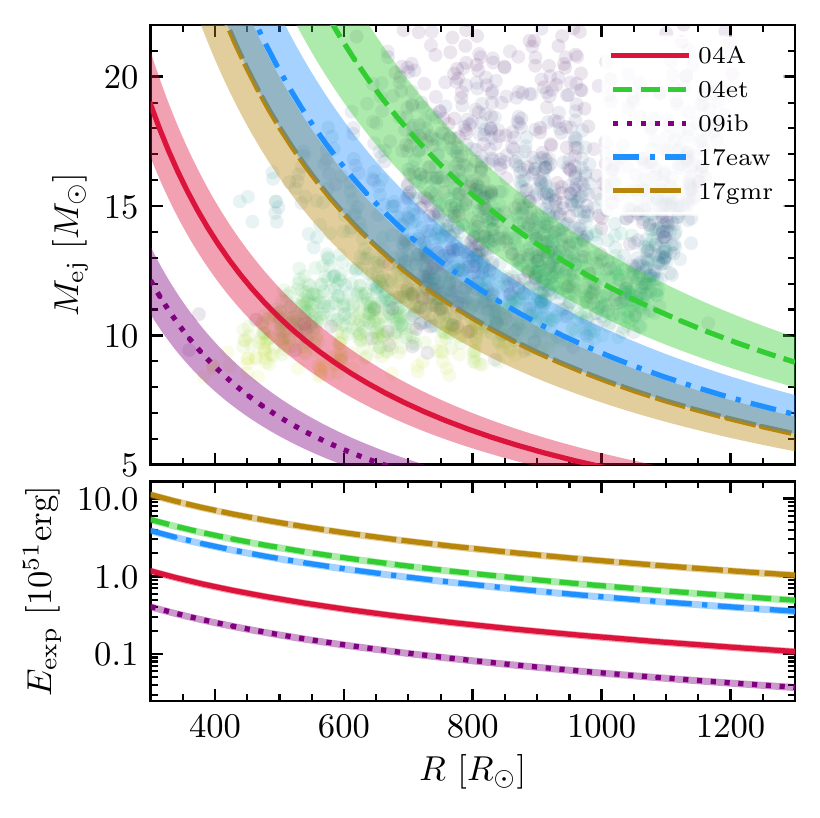} 
\caption{Degeneracy curves for $\Mej$ (top) and $\Eexp$ (bottom) recovered
from Equations~\eqref{eq:scaling} as a function of $R$
for the observed SNe considered here. 
Shaded solid-color regions correspond to the $\approx11\%$ RMS deviations between the models of GBP19 and their recovered 
parameters. Additional observational uncertainties are not included. The $\Mej$ and $R$ of 
2179 progenitor models are also shown in the background, with color ranging from yellow to 
purple tracking $\Mzams=10-25\Msun$.} 
\label{fig:wedge}
\end{figure}

The families of explosion parameters recovered by inserting each SN's $\MNi$, $\Lfifty$, and $\tpt$ 
into Equations~\eqref{eq:scaling} are shown in Figure~\ref{fig:wedge} as a function of $R$.
Also shown is a large suite of RSG progenitor models to demonstrate the potential 
variety of $\Mej$ and $R$ within reasonable stellar evolution assumptions. 
For each event, $\Mej$ and $\Eexp$ can be inferred from the plot for a given $R$. 

The progenitor models were constructed using \MESA\ revision 10398, 
and evolved to Fe core infall, following the \verb|example_make_pre_ccsn| test case described in detail by 
\citet[hereafter \mesafour]{Paxton2018}.
We varied the initial mass ($M_\mathrm{ZAMS}/\Msun=$10.0-15.0 in increments of 0.5$\Msun$, and 15.0-25.0 in increments of 1.0$\Msun$), 
surface rotation ($\ooc=0.0;\,0.2$), mixing length $\alpha$ in the H-rich envelope 
($\alpha_\mathrm{env}$=2.0; 3.0; 4.0), core overshooting 
($f_\mathrm{ov}=0.0$; 0.01; 0.018), 
and wind efficiency ($\eta_\mathrm{wind}=0.1-1.0$, increments of 0.1) using \mesa's `Dutch' wind 
scheme.
All models had $Z=0.02$. Only models which reached Fe core infall are shown. 
Rather than one relationship between $\Mej$ and $R$, this set of models suggests a wide range 
in which RSGs can exist. This diversity reflects the
importance of winds in determining the final masses and radii of stellar models \citep{Renzo2017}, 
and supports recent work showing diversity in progenitor masses for comparable positions 
on the HR diagram (\citealt{Farrell2020}).

\section{Explosion Models and Comparison to Observations\label{sec:observations}}
We then select progenitor models to explode in order to match observations
guided by Equations~\eqref{eq:scaling} applied to a SN's respective $\Lft$, $\MNi$, and $\tpt$. 
For SNe 2004A, 2004et, SN2017eaw, and 2017gmr, we chose three progenitor models each, consistent with the respective 
degeneracy curves in Figure~\ref{fig:wedge}, with ejecta masses near the larger-$\Mej$, middle-$\Mej$, 
and smaller-$\Mej$ intersections of the theoretical curves and the progenitor model suite.
For SN2017eaw, we chose three additional models consistent with a distance 10\% farther away 
(i.e. increasing $\Lfifty$ and $\MNi$ by 21\%, not shown in Figure~\ref{fig:wedge}). 
Very low $\Mej$ and radii are recovered for SN2009ib, with little overlap with our progenitor grid, 
so we exploded only two progenitors, one off the grid ($\alpha=6$). 
Properties of these models at the moment of explosion, input physics, and values for $\MNi$ 
are shown in Table \ref{tab:models}. Also shown are the time to shock breakout ($t_\mathrm{sb}$) and 
the mass above the photosphere at day 20 ($\delta m_\mathrm{ph,20}$).

\begin{table*}
\caption{Model Properties}
\centering
\singlespace
\small
\begin{tabular}{ccccccccccc}
\hline\hline
SN Name   & Model name                         & $M_\mathrm{ZAMS}$ & $f_\mathrm{ov}$, $\alpha_\mathrm{env}$, $\ooc$, $\eta_\mathrm{w}$  & $\Mf$     & $M_\mathrm{c,f}$  & $M_\mathrm{c, He}$ & log($\frac{L_p}{L_\odot}$)  & $\Teff$ & $t_\mathrm{sb}$  & $\delta m_\mathrm{ph,20}$   \\   
($\MNi/\Msun$) &[$M_\mathrm{ej,\odot}$][$R_\odot$][$\Efoe$]& [$\Msun$]             &                 & [$\Msun$] & [$\Msun$]       & [$\Msun$]       &                  & [K]  & [days] & [$M_\odot$] \\\hline    
2004A    &M9.3\_R596\_E0.4                    & 11.5           & 0.018, 3.0, 0.0, 0.5                & 10.87     & 1.62            & 3.79            & 4.86             & 3900      & 1.6 & 0.032 \\  
(0.042)  &M10.6\_R482\_E0.5                   & 12.5           & 0.01, 4.0, 0.0, 0.2                 & 12.28     & 1.64            & 3.89            & 5.20             & 5250      & 1.2 & 0.061  \\  
         &M15.2\_R438\_E0.8                   & 17.0           & 0.0, 4.0, 0.0, 0.2                  & 16.66     & 1.48            & 5.33            & 5.23             & 5610      & 1.0 & 0.096  \\  
\hline
2004et   &M11.8\_R945\_E0.76                  & 14.0           & 0.018, 2.0, 0.2, 0.2                & 13.42     & 1.59            & 4.89            & 5.22             & 3790      & 2.2 & 0.031  \\  
(0.063)  &M14.9\_R816\_E1.0                   & 18.0           & 0.0, 2.0, 0.0, 0.5                  & 16.53     & 1.62            & 5.85            & 5.44             & 4640      & 1.8 & 0.036  \\  
         &M18.3\_R791\_E1.2                   & 22.0           & 0.0, 3.0, 0.0, 0.5                  & 19.89     & 1.55            & 7.70            & 5.25             & 4160      & 1.7 & 0.040  \\  
\hline
2009ib   &M7.86\_R375\_E0.23                  & 10.0           & 0.018, 4.0, 0.2, 0.7                & 9.41      & 1.55            & 3.15            & 5.05             & 5450      & 1.1 & 0.074  \\ 
(0.043)  &M10.2\_R356\_E0.3                   & 12.0           & 0.01, 6.0, 0.2, 0.4                 & 11.65     & 1.48            & 3.69            & 3.99             & 3040      & 1.1 & 0.082  \\ 
\hline
2017eaw  &M10.2\_R850\_E0.65                  & 13.5           & 0.01, 2.0, 0.2, 0.8                 & 11.99     & 1.77            & 4.24            & 4.92             & 3370      & 2.0 & 0.032  \\  
at 6.85Mpc&M12.7\_R719\_E0.84                  & 15.0          & 0.01, 3.0, 0.0, 0.2                 & 14.53     & 1.80            & 5.09            & 5.04             & 3910     & 1.7 & 0.036  \\  
(0.048)   &M17.2\_R584\_E1.3                   & 20.0          & 0.0, 4.0, 0.0, 0.4                  & 18.92     & 1.70            & 6.79            & 5.10             & 4490     & 1.2 & 0.072  \\  
\hline
2017eaw, mod.  &M11.9\_R849\_E0.9              & 14.0          & 0.016, 2.0, 0.0, 0.2                & 13.64     & 1.70            & 4.55            & 5.08             & 3690     & 1.8 & 0.032 \\  
at 7.54Mpc&M15.7\_R800\_E1.1                   & 19.0          & 0.0, 3.0, 0.2, 0.4                  & 17.33     & 1.66            & 6.83            & 5.18             & 4040     & 1.7 & 0.041   \\ 
(0.0581)  &M19.0\_R636\_E1.5                   & 22.0          & 0.0, 4.0, 0.0, 0.2                  & 20.51     & 1.55            & 7.74            & 5.54             & 5550     & 1.2 & 0.056   \\ 
\hline
 2017gmr &M9.5\_R907\_E1.9                    & 12.0           & 0.018, 2.0, 0.2, 0.6                & 11.01     & 1.48            & 3.86            & 5.70             & 5110      & 1.1 & 0.076  \\  
 (0.13)  &M12.5\_R683\_E3.0                   & 14.5           & 0.01, 3.0, 0.0, 0.2                 & 14.09     & 1.55            & 4.80            & 5.46             & 5120      & 0.81 & 0.11  \\  
         &M16.5\_R533\_E4.6                   & 19.0           & 0.0, 4.0, 0.0, 0.4                  & 18.09     & 1.57            & 6.28            & 5.29             & 5250      & 0.55 & 0.22  \\  
\hline          
\label{tab:models} 
\end{tabular} 
\end{table*}

We then excised the Fe cores with an entropy cut of 4 erg g$^{-1}$ K$^{-1}$, and exploded 
these models using \MESA\ with Duffell RTI \citep{Duffell2016} and the fallback estimation technique described in 
Appendix A of GBP19, with an additional velocity cut of 500 km s$^{-1}$ at handoff to \stella\ at shock 
breakout.\footnote{For all models except 2017eaw at 6.85~Mpc, 
\MESA\ revision 10925 was used, as in GBP19. Because we consider excess emission in the early lightcurve of 
2017eaw at 6.85~Mpc, revision 11701 was used with a dense mesh near the surface set by 
`\texttt{split\_merge\_amr\_logtau\_zoning=.true.}' in \texttt{inlist\_controls}
to ensure that the outer region 
is adequately resolved.} All explosions resulted in negligible fallback. 
At shock breakout, we rescaled the $\Ni$ distribution to match the desired $\MNi$, 
and imported the ejecta profile into \STELLA\ to model the evolution post-shock-breakout.  
We used 400 spatial zones and 40 frequency bins in \STELLA, which yields
convergence in bolometric lightcurves on the plateau (see Figure~30 
of \mesafour\ and the surrounding discussion). For SN2017eaw at 6.85~Mpc, we used 800 spatial zones in order to more 
faithfully capture the outermost layers of the ejecta. 
Because we are focused on matching plateau emission from the bulk of the ejecta, occurring after day $\approx20$, 
we do not include any extra material beyond the progenitor photosphere for most of our model lightcurves. 

\subsection{Comparison to Observed SNe\label{sec:normal}}
\begin{figure}
\centering
\includegraphics{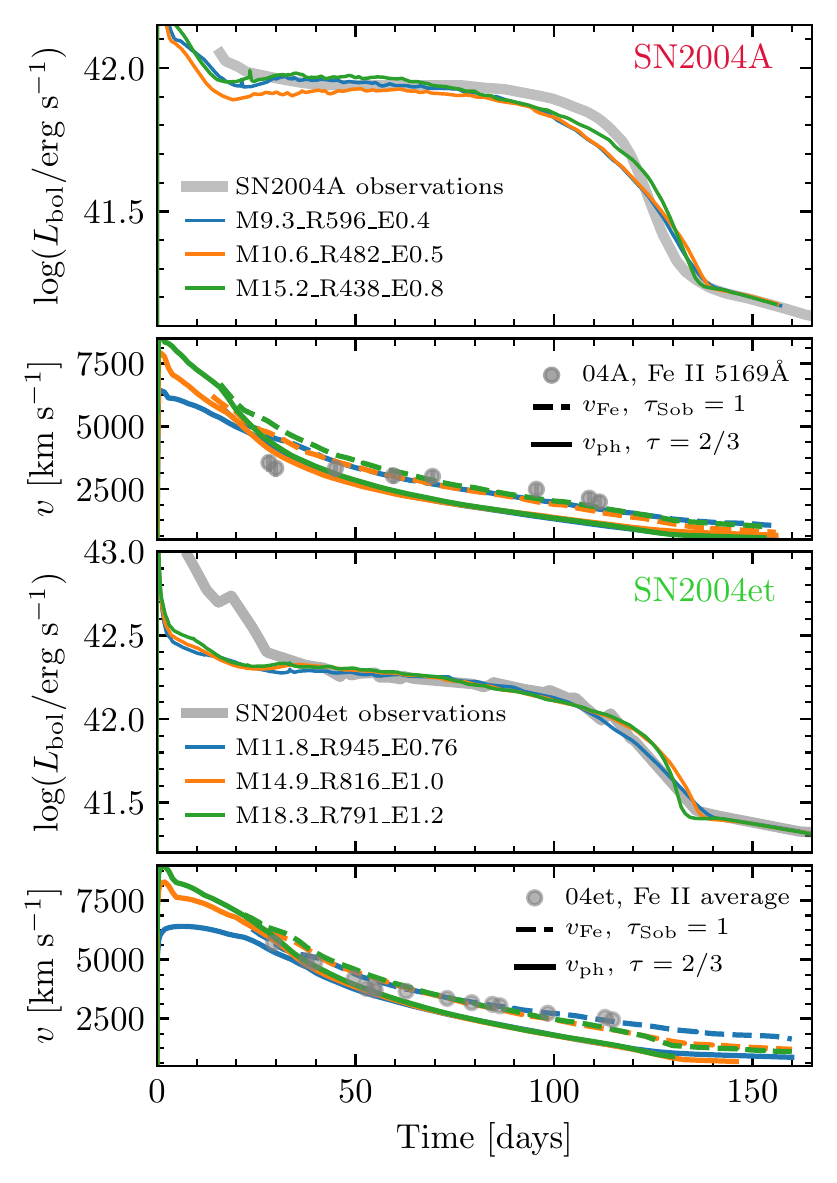} 
\caption{Lightcurves and Fe line velocities for SN2004A (top two panels) 
and SN2004et (bottom two panels). Grey markers correspond to the observations, 
and colored lines correspond to explosion models, ordered in ascending
$\Mej$ and $\Eexp$, and descending $R$.} 
\label{fig:SN2004A}
\end{figure}

Despite intrinsic scatter amounting to
$\approx$11\% RMS deviations between model parameters and $\Mej$ and $\Eexp$ recovered 
from Equations~\eqref{eq:scaling} applied to model radii and lightcurves (GBP19), 
computations approximately obeying Equations~\eqref{eq:scaling} produce bolometric lightcurves 
which match the observations. Figure~\ref{fig:SN2004A} shows 
the results for SN2004A (top two panels) and SN2004et (bottom two panels). Both SN2004A and 
SN2004et exhibit good agreement between models, lightcurves, 
and velocity evolution on the plateau, with no model being the ``best-fit" for either event. 
Photospheric velocities at very early times ($\ltapprox 20$ days) do differ between different models, 
with more compact, higher-$\Eexp$ models yielding faster early-time velocities. However,
velocity measurements before day 20 are rare, and at these times velocities might be modified by the 
circumstellar environment (e.g. \citealt{Moriya2018}). 
The early observed lightcurve ($\ltapprox30$ days) of SN2004et also exhibits 
a clear luminosity excess compared to the lightcurve models. Such excess is often attributed to 
interaction with an extended envelope or wind, or with pre-SN outbursts\citep[e.g.][]{Morozova2017,Morozova2020}.

All three models for SN2004et are consistent with the reported $R=350-980\Rsun$. 
For SN2004A, only the low-mass/low-energy 
model M9.3\_R596\_E0.4 is consistent with the progenitor observations, 
and we conclude for that SN that $\Mej\ltapprox10\Msun$ and
$\Eexp\ltapprox0.4\times10^{51}\,\mathrm{erg}$.

\subsection{SN2017eaw at Two Distances\label{sec:17eaw}}
\begin{figure}
\centering
\includegraphics{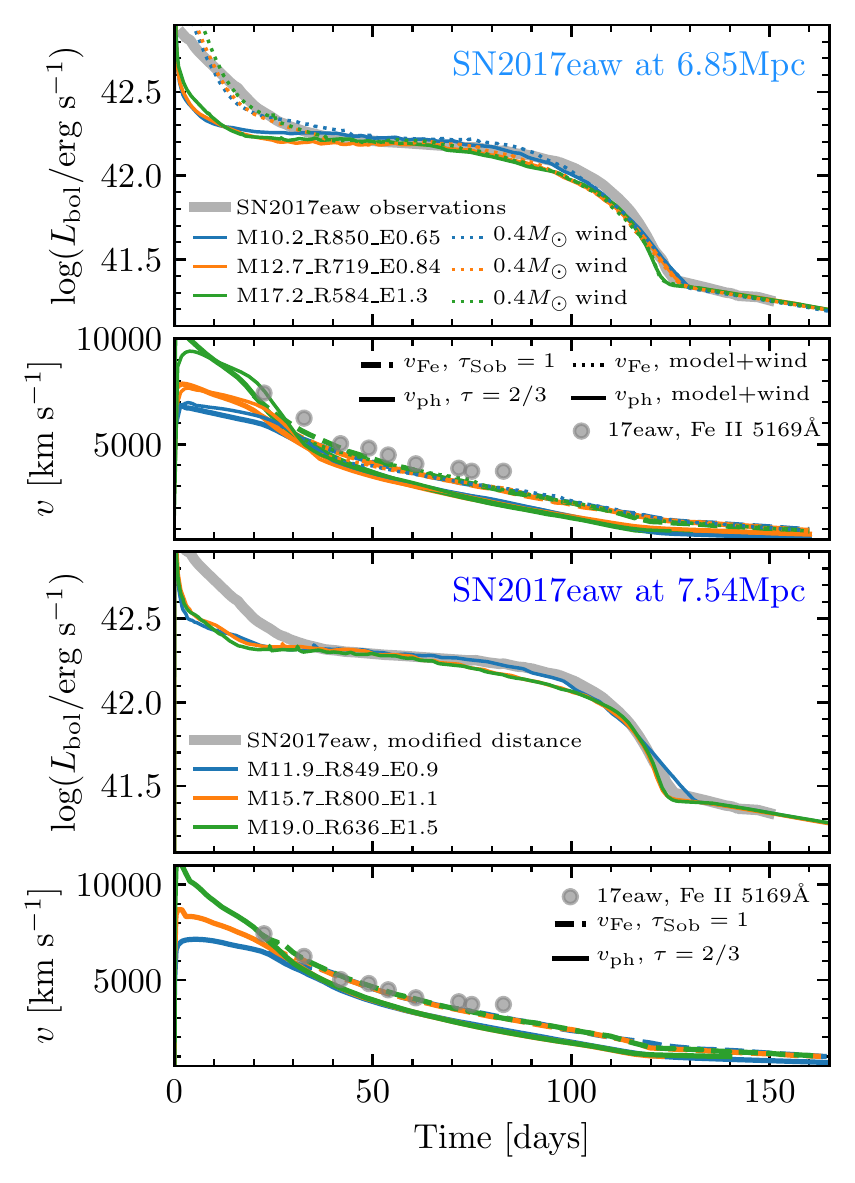} 
\caption{Lightcurves and Fe-line velocities for observations and models of 
SN2017eaw at $D=$6.85~Mpc (top two panels) 
and $D=$7.54~Mpc (bottom two panels). Grey markers correspond to observations, 
and colored lines correspond to explosion models. 
Colored dotted lines in the upper panels correspond to models with an additional 0.4$\Msun$ wind 
($v_\mathrm{wind}=8\,\mathrm{km\,s^{-1}}$, $\dot{M}_\mathrm{wind}=0.2\,\Msun$/year).} 
\label{fig:2017eaw}
\end{figure}

To show the impact of changing the assumed distance on our modeling, 
we model SN2017eaw at two different distances: 
6.85~Mpc, using the fiducial \citet{Szalai2019} lightcurve, and at 7.54~Mpc, 
with the same $\tpt$ but with 21\% brighter $\Lbol$ and $\MNi=0.0581\Msun$. 
Models were selected to match Equations~\eqref{eq:scaling} with the appropriate $\Lft$, $\tpt$, and $\MNi$
for each distance. 
Figure~\ref{fig:2017eaw} compares models to observations.
The top two panels correspond to $D=6.85~\mathrm{Mpc}$, 
and the bottom two panels to $D=7.54~\mathrm{Mpc}$.\footnote{The farther distance was motivated by the 
fact that velocities of models matching $\Lfifty$ and $\tpt$ of the fiducial distance are $\approx10\%$ discrepant with 
observed velocities. Since $\Lfifty\propto D^2\appropto\vFef^2$ (\citealt{Hamuy2002,Kasen2009}; GBP19), 
an intrinsically brighter SN at a distance $\approx10\%$ farther produces models which better match the velocity data.
This distance is also consistent with a recent TRGB estimate of $7.72\pm0.78~\mathrm{Mpc}$
\citep{VanDyk2019}.}
Like SN2004A and SN2004et, models agree well with the data, 
and agreement in $\Lfifty$ also yields agreement in the velocity of the models after day $\approx20$. 
Agreement between models and \textit{both} velocity and luminosity data is better for $D=7.54~\mathrm{Mpc}$. 
For $D=6.85~\mathrm{Mpc}$, two of our models, 
M10.2\_R850\_E0.65 and M12.7\_R719\_E0.84, match the progenitor properties within 
the uncertainties. At a $10\%$ farther distance, assuming
$21\%$ brighter $L_\mathrm{p}$ and the same $\Teff$, 
only our M11.9\_R849\_E0.9 model is consistent with the updated progenitor properties.
Assuming the measured progenitor radius of 
$845\Rsun$, we chose models with $R\approx850\Rsun$ for both distances.
The 10\% greater distance leads to $\approx17\%$ increase in $\Mej$, 
from $10.2\Msun$ to $11.9\Msun$ and $\approx40\%$ increase in $\Eexp$, 
from 0.65$\times\foe$ to 0.9$\times\foe$. 

For $D=6.85~\mathrm{Mpc}$, we also show lightcurves with and without a dense 
wind to reproduce the early excess emission (top two panels of Figure~\ref{fig:2017eaw}). 
We affix a wind density profile with total mass $M_\mathrm{wind}$ and
$\rho_\mathrm{wind}(r)=\dot{M}_\mathrm{wind}/4\pi r^2 v_\mathrm{wind}$,
where 
$\dot{M}_\mathrm{wind}$ is a constant, and $v_\mathrm{wind}$ is the wind velocity. 
We varied $\dot{M}_{\mathrm{wind}}=(0.1,0.2,0.3,0.4)\,\Msun/\mathrm{yr}$ 
and $v_\mathrm{wind}=(3,5,8,12)\,\mathrm{km/s}$ with $M_\mathrm{wind}$ from $0.2-0.8\,\Msun$.
In the top of Figure~\ref{fig:2017eaw} we show values of $v_\mathrm{wind}=8\,\mathrm{km\,s^{-1}}$, 
$\dot{M}_\mathrm{wind}=0.2\,\Msun$/year, and $M_\mathrm{wind}=0.4\Msun$.
We find that the same wind parameters produce comparable early excesses when added to the three degenerate lightcurves, 
suggesting that the excess is set by properties of the wind itself and the underlying lightcurve, 
rather than, e.g. $\Eexp$. This wind also modifies the early velocity evolution.
We do not claim that this is the only way to reproduce the early excess, 
as a variety of other outer density profiles can give rise to similar early 
excesses without affecting plateau properties \citep[e.g.][]{Morozova2020}.

\subsection{Modeling Challenges\label{sec:weird}} 
\begin{figure}
\centering
\includegraphics{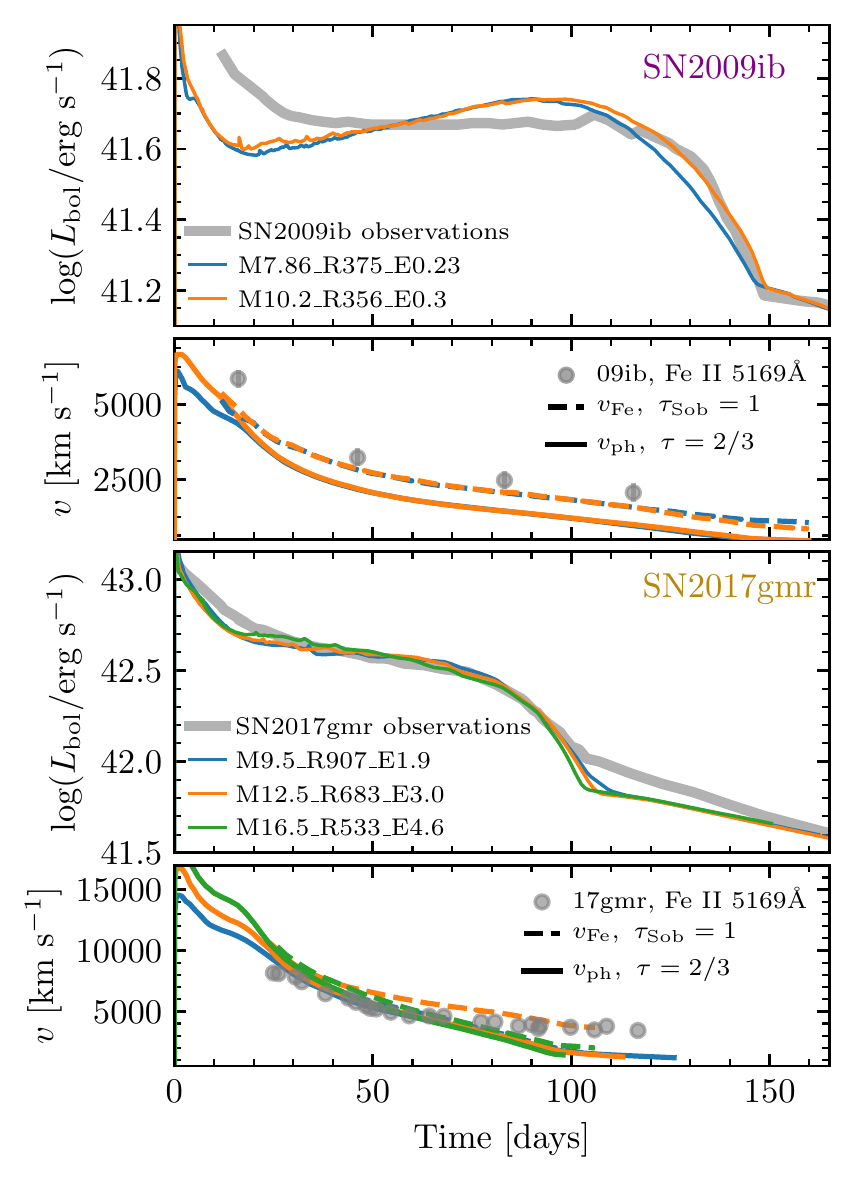} 
\caption{Lightcurves and Fe line velocities for SN2009ib (top two panels) 
and SN2017gmr (bottom two panels). Grey markers correspond to observations, 
and colored lines correspond to explosion models.} 
\label{fig:SN2009ib}
\end{figure}

For two events, SN2009IB and
SN2017gmr, we see general agreement between models and bulk properties of  the
lightcurves ($\Lfifty$ and $\tpt$), with distinct differences shown in 
Figure~\ref{fig:SN2009ib}. Specifically, these models differ beyond an early
luminosity excess which might be explained by pulsations, a wind, varied
structure of  the extended stellar atmosphere, or other early interaction. 

In SN2009ib (top two panels of Figure~\ref{fig:SN2009ib}), the relatively low luminosity and high 
$\Ni$ heating yield lightcurve models which rise significantly between days $20-80$. 
The narrow overlap between Equations~\eqref{eq:scaling} and our model grid suggests low $\Mej$ and small $R$. 
For a reasonable range of $R$, explosion energies recovered are also low ($\Efoe\approx0.2-0.3$).
The disagreement between the models and the velocity data may indicate that 
$\Lbol$ is underestimated in some 
way (see the discussion in section \ref{sec:17eaw}). 
However, we found that additional models consistent with the velocity data and a brighter lightcurve of 
the same $\tpt$ still exhibit a similar, slightly shallower positive plateau slope. 
It is also possible that the estimated explosion epoch is too early. 
Moreover, neither explosion is consistent with a RSG of
$R\approx1000\Rsun$ (derived assuming $\Teff=3400\mathrm{K}$), 
as $R=1000\Rsun$ implies exceedingly low $\Mej\approx3\Msun$ and
$\Eexp\approx6\times10^{49}\mathrm{erg}$. 
However, model M7.86\_R375\_E.023 is able to reproduce 
the observed $\log(L_\mathrm{p}/\Lsun)=5.04\pm0.2$ with 
$\Teff\approx5450\mathrm{K}$, thus favoring the yellower source scenario.

For SN2017gmr (bottom two panels of Figure~\ref{fig:SN2009ib}), all models agree equally well with the 
lightcurve, but indicate an excess in observed luminosity after the plateau 
as the lightcurve transitions to the $\Co$-decay tail. Observed velocities
are taken from the reported Fe-line radius evolution, and are only shown before day 120, after which point the 
evolution is not photospheric. 
The slight disagreement between modeled and observed velocities suggests that perhaps the distance is overestimated, 
but modeling to match a fainter bolometric lightcurve provides no change in the apparent late-time excess. 

Although this event has no progenitor pre-image,
if $R$ at the time of explosion is consistent with 
$\approx500\Rsun$ recovered from fitting shock-cooling models to 
the photometric bands \citep{Andrews2019}, Equations~\eqref{eq:scaling} imply 
an enormous $\Eexp\approx5\times10^{51}$ ergs! Our $533\Rsun$ progenitor model indeed 
matches $\Lfifty$ and $\tpt$ when exploded with $4.6\times\foe$, 
shown in green in the lower two panels of Figure~\ref{fig:SN2009ib}.

Our modeling procedure only considers matching $\Lfifty$ and $\tpt$.
To compare directly to the day~1 results in \citet{Andrews2019} Figure 9,
we re-ran the SN2017gmr models with a surface resolution adequate to resolve emission at 
day~1 ($\delta m_\mathrm{ph}\sim10^{-3}-10^{-4}\Msun$). 
All three of our models yield luminosities at 1~day post-shock-breakout ($L_1$)
a factor of $\approx2$ lower than $L_1$ of SN2017gmr recovered by
their \citet{Sapir2017} shock-cooling model fits. 
Of our models, the day~1 photospheric temperature ($T_1$) of M16.5\_R533\_E4.6 
does come closest to the reported shock-cooling $T_1=25,900\mathrm{K}$, 
with $T_1\approx27,000\mathrm{K}$, 
as compared to 29,000K for M12.5\_R683\_E3.0 and 30,000K for M9.5\_R907\_E1.9.
At this time in the lightcurve evolution, the emitting region is 
coincident with the location of a density inversion in the stellar models, 
which is the focus of current ongoing studies.

For the lightcurve morphological differences, we have no easily available remedy without additional free parameters. 
Because we use the \citet{Duffell2016} mixing prescription with coefficients calibrated to the 3D 
simulations as recommended in \mesafour, the resulting smoothing of the density profile and compositional mixing
are held `fixed.' Nonetheless, the Equations~\eqref{eq:scaling}-motivated models agree well with 
the $\Lfifty$ and $\tpt$ observations. 

\section{Conclusions}
The capability of \MESA+\STELLA\ to model observed SNe was introduced in \mesafour\ and demonstrated there and 
by \citet{Ricks2019} to model a few Type IIP SNe. 
GBP19 introduced scaling relations (Equations~\ref{eq:scaling}) fit from a suite of \MESA+\STELLA\ models
in order to guide explosion modeling efforts for an observed SN lightcurve with a given $\Lfifty$, $\tpt$, and $\MNi$. 
In the absence of understanding in models of the first 20 days,
our application of these relations to the observed SNe 2004A, 2004et, 2009ib, 2017eaw, and 2017gmr 
shows families of explosion models that match the data for a wide range of $\Mej$, $R$, and $\Eexp$.
These degeneracies will not be easily lifted without an observed progenitor radius 
(and understanding the progenitor's variability; see \citealt{Goldberg2020}) or other constraints. 
However, when combined with a radius given by progenitor pre-imaging or fitting 
the shock-cooling phase, we show that explosion models can be constrained following 
$\Eexp\propto\,R^{-1.63}$ and $\Mej\propto\,R^{-1.12}$.

If there was confidence in stellar evolutionary input constraining a $R-\Mej$ 
relation at the time of explosion, then these degeneracies could be broken, 
as assumed in the population synthesis/lightcurve 
modeling of \citet{Eldridge2019}. However, when varying rotation, winds, 
core overshooting, and mixing length within a reasonable 
range of values, we find no single ejecta-mass$-$radius relation.

It remains possible that detailed spectral modeling will lend insights 
which might aid in uniquely determining explosion properties from plateau observations. 
Additionally, velocity observations before day $\approx20$ 
or photospheric radii derived from shock-cooling models
with a secure density structure in the outer $<0.1\,\Msun$
remain other promising paths forward to breaking the remaining degeneracies 
exhibited here.

\acknowledgements
We thank Bill Paxton for continued support and advancement of 
\MESA's capabilities, and Josiah Schwab and Benny Tsang for 
conversations and guidance. 
We thank the referees for helpful comments that significantly 
improved our presentation.
We thank J\'ozsef Vink\'o and 
Tam\'as Szalai for providing bolometric data for SN2017eaw. 
It is a pleasure also to thank 
K. Azalee Boestrom, 
Daichi Hiramatsu,
D. Andrew Howell,
and
Stefano Valenti
for correspondences about observations.

J.A.G. is supported by the National Science Foundation (NSF) GRFP grant No.\,1650114. 
The \MESA\ project is supported by the NSF under the Software Infrastructure for Sustained 
Innovation program grant ACI-1663688. This research was supported in part by the Gordon 
and Betty Moore Foundation through Grant GBMF5076 and at the KITP by the 
NSF under grant PHY-1748958. We acknowledge the use of computational 
facilities through the Center for Scientific Computing at the CNSI,
MRL: an NSF MRSEC (DMR-1720256) and NSF CNS-1725797.

This research made extensive use of the SAO/NASA Astrophysics 
Data System (ADS).

\software{
\texttt{MESA}, 
\texttt{STELLA}, 
\texttt{py\_mesa\_reader} \citep{MesaReader}, 
\texttt{SciPy} \citep{scipy}, 
\texttt{matplotlib} \citep{hunter_2007_aa}.
}

\bibliographystyle{yahapj}


\end{document}